\documentclass[nofootinbib, twocolumn, amsmath,amssymb, aps,superscriptaddress]{revtex4-2}

\usepackage{graphicx}
\usepackage{dcolumn}
\usepackage{bm}
\usepackage{hyperref}

\usepackage{color}
\usepackage[normalem]{ulem}
\usepackage[dvipsnames]{xcolor}
\usepackage{placeins}

\hypersetup{colorlinks=true, citecolor=MidnightBlue,linkcolor=CornflowerBlue, urlcolor=CornflowerBlue, linktocpage=true}

\newcommand{\mpl}{M_{\rm Pl}}

\begin{document}

\preprint{APS/123-QED}

\title{Testing gravitational wave propagation with multiband detections}
 \author{Tessa Baker}
  \email{t.baker@qmul.ac.uk}
 \affiliation{Department of Physics \& Astronomy, Queen Mary University of London, Mile End Road, London, E1 4NS, UK}
  \author{Enrico Barausse}
 \email{barausse@sissa.it}
 \affiliation{
  SISSA - Scuola Internazionale Superiore di Studi Avanzati, Via Bonomea 265, 34136 Trieste, Italy and INFN Sezione di Trieste \\
  IFPU - Institute for Fundamental Physics of the Universe, Via Beirut 2, 34014 Trieste, Italy \\
 }
\author{Anson Chen}
\email[corresponding author -- ]{a.chen@qmul.ac.uk}
\affiliation{Queen Mary University of London, Mile End Road, London, E1 4NS, UK}
\author{Claudia de Rham}
\email{c.de-rham@imperial.ac.uk}
 \affiliation{Theoretical Physics, Blackett Laboratory, Imperial College, London, SW7 2AZ, UK}
 \affiliation{Perimeter Institute for Theoretical Physics,
31 Caroline St N, Waterloo, Ontario, N2L 6B9, Canada}
\author{Mauro Pieroni}
\email[corresponding author -- ]{m.pieroni@imperial.ac.uk}
 \affiliation{Theoretical Physics, Blackett Laboratory, Imperial College, London, SW7 2AZ, UK}
 \author{Gianmassimo Tasinato}
\email{g.tasinato@swansea.ac.uk}
\affiliation{Dipartimento di Fisica e Astronomia, Universit\`a di Bologna, via Irnerio 46, Bologna,  Italy}
 \affiliation{Physics Department, Swansea University, SA2 8PP, UK}

\date{\today}

\begin{abstract}
 \noindent Effective field theories (EFT) of dark energy (DE) -- built to parameterise the properties of DE in an agnostic manner -- are severely constrained by measurements of the propagation speed of gravitational waves (GW). However, GW frequencies probed by ground-based interferometers lie around the typical strong coupling scale of the EFT, and it is likely that the effective description breaks down before even reaching that scale. We discuss how this leaves the possibility that an appropriate ultraviolet completion of DE scenarios, valid at scales beyond an EFT description, can avoid present constraints on the GW speed. Instead, additional constraints in the lower frequency LISA band would be harder to escape, since the energies involved are orders of magnitude lower. By implementing a method based on GW multiband detections, we show indeed that a single joint observation of a GW150914-like event by LISA and a terrestrial interferometer would allow one to constrain the speed of light and gravitons to match to within $10^{-15}$. Multiband GW observations can therefore firmly constrain scenarios based on the EFT of DE, in a robust and unambiguous way.
\end{abstract}
\maketitle

\section {Introduction}
\label{sec:intro}
Identifying the  source of the current accelerated expansion of the Universe represents one of the biggest challenges in cosmology today. Models ascribing cosmological observations to a cosmological constant or to a dark energy (DE) component have been proposed, but are hardly natural from a microscopic point of view~\cite{Weinberg:1988cp}.  Given that all evidence for  cosmological acceleration is gravitational in nature, it is possible that cosmological data can  be explained without any such components, if deviations from General Relativity (GR) occur in the low energy  regime, relevant for cosmology.

The simplest option for modifying GR is to include an extra  scalar degree of freedom,  non-minimally coupled with gravity, and whose time-dependent profile causes cosmic acceleration with no need of a cosmological constant. Over the past decades several scalar-tensor theories have been proposed, with increasing degrees of generality, from Fierz-Jordan-Brans-Dicke~\cite{Brans:1961sx} theory, to Horndeski~\cite{Horndeski:1974wa}, beyond Horndeski~\cite{Zumalacarregui:2013pma,Gleyzes:2014dya}, to DHOST~\cite{Langlois:2015cwa,BenAchour:2016fzp,Crisostomi:2016czh}. A powerful theoretical framework, the effective field theory (EFT) of DE (see e.g.~\cite{Gleyzes:2013ooa,Gleyzes:2014rba,Lagos:2016wyv,Lagos:2017hdr}) allows one to encapsulate  the predictions of such scalar-tensor theories in terms of a set  of  operators, describing the possible couplings of the scalar field with gravity. Such an approach is  useful for investigating the regime of  validity of a theory and its energy cutoff, and for confronting scalar-tensor theories with cosmological data in terms of a small number of parameters: see~\cite{Joyce:2014kja} for a comprehensive review. 

The operators appearing in the EFT of DE generically include non-minimal, derivative couplings of the metric to the scalar field. Once the scalar acquires a  time-dependent profile -- as expected if it drives cosmic self-acceleration -- the resulting speed of gravitational waves (GW) is  different from the speed of light.

Using this fact, significant constraints on scalar-tensor theories have been placed by the detection of GWs in the LIGO-Virgo-KAGRA (LVK) band. In particular,  the measured  times of arrival of the GW signal from the neutron star (NS) merger GW170817, and of the electromagnetic signal from GRB 170817A, were coincident within two seconds~\cite{LIGOScientific:2017vwq,LIGOScientific:2017zic,LIGOScientific:2017ync}. This observation allowed the relative speeds of GW and gamma rays to be constrained to within a fractional difference $\lesssim 10^{-15}$~\cite{LIGOScientific:2017vwq}, which in turn sets tight bounds on existing gravity theories~\cite{Lombriser:2015sxa,Joyce:2016vqv, Lombriser:2016yzn,Creminelli:2017sry,Sakstein:2017xjx,Ezquiaga:2017ekz,PhysRevD.95.084029,Baker:2017hug,Kubota:2022lbn}. However, the frequency of the GW170817 signal is  expected to be above the cutoff in these EFTs of DE. Hence, it may be natural for the speed of GWs to be luminal at LVK frequencies while still departing from luminality at much lower frequencies~\cite{deRham:2018red}.

Thanks to its lower frequencies of operation (in the mHz band), the future space-borne Laser Interferometer Space Antenna (LISA)~\cite{LISA:2017pwj} will provide GW observations at  lower energies than the EFT strong coupling scale, thus potentially constraining scalar-tensor theories in a more robust and unambiguous way. However, the existence of electromagnetic counterparts for LISA is highly uncertain~\cite{Tamanini:2016zlh,LISACosmologyWorkingGroup:2019mwx}; hence bounding the propagation speed of GWs with the massive black hole equivalent of standard sirens may not be feasible in practice.

As we discuss in this work, a promising alternative may instead be provided by multiband observations of stellar-origin black hole binaries (SOBHBs) with LISA and LVK (or a third generation terrestial GW detector). In fact, Ref.~\cite{Sesana:2016ljz} shows that GW signals from this family of binaries, already detected by LVK, will be also observable with LISA in their early-stage, low frequency inspiral.  LISA observations will take place months to years prior to the final merger in the band of terrestrial detectors, allowing for the merger time to be predicted to within tens of seconds~\footnote{These results have been confirmed by full Bayesian analyses~\cite{Toubiana:2020vtf,Toubiana:2020cqv}, at least for quasi-circular binaries. More recently, Ref.~\cite{Klein:2022rbf} found much larger uncertainties on the predicted merger time than Refs.~\cite{Toubiana:2020vtf,Toubiana:2020cqv}. However, the disagreement disappears~\cite{toubiana_private} for quasi-circular binaries, once a higher-PN order (and thus more accurate) expression for the merger time is employed in the analysis of Ref.~\cite{Klein:2022rbf}. 
 }. For this reason, Refs.~\cite{Barausse:2016eii,Toubiana:2020vtf} proposed to use multiband observations of SOBHBs as a test  for the presence of non-GR gravitons, which would lead to additional GW emission from the binary (on top of the usual tensor emission described at leading order by the quadrupole formula), and thus to an earlier merger. Additionally,~\cite{Vitale:2016rfr,Carson:2019rda,Gnocchi:2019jzp} showed that multiband GW observations can improve theory-agnostic bounds on deviations from GR in the inspiral as well as  test  specific alternative theories of gravity, while~\cite{Gupta:2020lxa,Datta:2020vcj} emphasized how multiband detections allow for breaking  degeneracies and performing  multiparameter tests of theories beyond GR.

In this paper, we show that a SOBHB merger event, which occurs  in the band of terrestrial detectors at a time different than that predicted from LISA observations, may also be caused by a non-linear graviton dispersion relation. We  assume a scenario  whereby GW generation is described by GR (possibly thanks to non-linear ``screening mechanisms''~\cite{deRham:2012fw,deRham:2012fg,Dar:2018dra,terHaar:2020xxb,Bezares:2021dma,Bezares:2021yek}, see e.g.~\cite{Burrage:2017qrf,Babichev:2013usa} for reviews) and only GW propagation is affected by modified gravity. In particular, we  make the hypothesis that the propagation speed of GW is $c_T=c_{\rm EM}$ (where the EM stands for Electro Magnetic) in the  band of ground-based experiments (as a result of new physics near the strong coupling scale of the EFT), while $c_{\rm LISA}=c_{\rm EM} -\delta c$ in the LISA band, with $\delta c/c_{\rm EM}$ very small. Then the LISA GW emission would be described by the same strain signal $h$ as in GR, but evaluated at the retarded time $t-D/c_{\rm LISA}$ (where $D$ is the distance of the source; notice  that since SOBHBs are only observable at low $z\lesssim 0.1$, cosmological effects on time and distance can be safely neglected). A LISA observer would then predict  the merger to  occur  at a time $t_{\rm LISA}=t_p+D/c_{\rm LISA}\pm \sigma_t$, where $t_p$ is the time at which the strain amplitude $h(t)$ is the largest (which corresponds to  coalescence), while  $\sigma_t\sim 5$ sec is the statistical error on the merger time estimate coming from the LISA data analysis. However, when later detected in the LVK band, the merger actually happens  at $t_{\rm Earth}=t_p+D/c_{\rm EM}$, i.e. at a different time than  predicted by  LISA. Since the merger time in ground-based detectors can be measured with a precision of  milliseconds, it is possible  to constrain 
\begin{equation}
|t_{\rm Earth}-t_{\rm LISA}| \approx \frac{D}{c_{\rm EM}} \,\left|\frac{\delta c}{c_{\rm EM}}\right| \,\lesssim \, \sigma_t\,,
\label{eq:t_delta_c}
\end{equation}
where, as mentioned above, we  assume $\delta c/c_{\rm EM}$ very small. For the distance of the first observed GW event, GW150914, $D\approx 410$ Mpc and this bound becomes $|\delta c|/c_{\rm EM}\lesssim  10^{-16}$ at $1\sigma$ confidence level (see Appendix~\ref{app_const} for more details). In this paper we  explore this idea, applying a Fisher Information Matrix (FIM) forecast to LISA data analysis, and showing that the  estimate of Eq.~\eqref{eq:t_delta_c} holds. Multiband observations of SOBHBs with LISA can therefore severely constrain (or confirm!) scenarios based on the EFT of DE, in a more robust and unambiguous manner than ground-based observations alone. First, though, we start with explaining our theoretical motivations for a frequency-dependent speed of GW.

During the last parts of this analysis, we became aware of similar methods and results being derived in~\cite{Harry:2022zey}. The precise details of that analysis differ from the current work in places: for example, the authors of~\cite{Harry:2022zey} consider a `distinguishability criterion' to assess if the time-shift of their waveforms induced by varying $c_T$ is recognisable. In the present work we instead perform a FIM analysis, which allows us to make statements about simultaneous constraints on regular and modified gravity parameters, and their degeneracies. The conclusions presented here are nevertheless in perfect agreement with those of~\cite{Harry:2022zey}.

We work in units where $\hbar=1$ and where the absolute speed of massless and free fields in the vacuum is $c=1$. Note however that we keep the speed of light denoted by $c_{\rm EM}$, as in principle this could be slightly different. The Planck scale $\mpl$ is related to Newton's constant by $\mpl=(8\pi G)^{-1/2}$.

\section{Theoretical Motivations}
\label{sec:theory}
For sake of concreteness, we present an example of a low-energy EFT, where the speed $c_T$ of GWs is a function of frequency. We emphasize however that the methods and conclusions derived in this work apply well beyond the scope of this illustrative example. As a warm-up,  we  first consider the  EFT for a single scalar field $\phi$ living in flat spacetime with low-energy interactions of the form
\begin{equation}
\mathcal{L}=-\frac 12 (\partial \phi)^2-\frac1{2\Lambda^4}(\partial \phi)^2\mathcal{G}\left(\frac{\Box}{\Lambda^2}\right)(\partial \phi)^2\,,
\end{equation}
where $\mathcal{G}$ a  function depending on the precise details of the  high-energy (UV) completion, and  $\Lambda$ the cutoff.
This low-energy EFT may be obtained by integrating out massive modes $H$ of mass above $\Lambda$, with interactions involving $H(\partial \phi)^2$ as described in~\cite{deRham:2018red}. We are not interested in the precise details of the UV completion here, but we expect the UV limit to be insensitive to the spontaneous breaking of Lorentz invariance induced at low energies by a scalar background. In fact, as mentioned  above, if $\phi$ were to play the role of DE field  causing self-acceleration, it would acquire a time-like profile. We can parameterize it as
\begin{equation}
\label{eq:scalpr}
\langle \phi \rangle \sim \alpha \Lambda^2 t \,,
\end{equation}
where $t$ the physical time, and $\alpha$ a dimensionless constant that may be used to relate the scale of the background with that of the low-energy EFT.
Fluctuations  on top of the background scalar in eq.~\eqref{eq:scalpr} are characterized by a frequency-dependent phase velocity $c_s(k)$, given by
\begin{equation}
c_s^2(k)=1+\alpha^2\,c_s^2(k)\,\mathcal{G}
\left(\left(c_s^2(k)-1\right)\frac{k^2}{\Lambda^2}\right) \,.
\end{equation}
In~\cite{deRham:2018red} a power-law profile for ${\mathcal G}$ was considered. However, in principle the function $\mathcal{G}$ can potentially be  sharper, and describe an abrupt transition between a constant non-luminal low-energy speed at low-energy $c_s(k\ll \Lambda)\sim 1-\delta c$, and a luminal speed at higher frequencies $c_s(k\gg \Lambda)=1$.

\smallskip
Turning our attention back to GWs, derivative mixing terms between gravity and other light degrees of freedom  are motivated by the EFT of DE, as explained in Section~\ref{sec:intro}. For instance, in the presence of a light scalar field $\phi$, the Einstein-Hilbert action may be supplemented with non-minimal mixings while the EM sector remains minimally coupled to gravity, 
\begin{equation}
\mathcal L=\frac{\mpl^2}{2}R-\frac 12 \mathcal{G}_{\mu\nu}\nabla^\mu\phi \nabla^\nu \phi-\frac 14 F_{\mu\nu}^2\,,
\end{equation}
where generic dark-energy EFT  operators are included within the quantity $\mathcal{G}_{\mu\nu}$ given by
\begin{equation}
\label{defGmn1}
 \mathcal{G}_{\mu\nu}=g_{\mu\nu}+\sum_{\ell, m,n}\frac{\nabla^m}{\Lambda^m}\frac{ \mpl^n R^n}{\Lambda^{3n}}\frac{\phi^\ell}{\Lambda^\ell}\,. 
\end{equation}
In principle one could also expect the EFT to include pure-potential terms but those are irrelevant to the current analysis and would be absent or suppressed if the scalar is endowed by a shift symmetry.
In the previous equation, we can in principle consider any covariant contraction of the Riemann tensor $R$, with an arbitrary number of higher derivatives. It is convenient, though,  to start the expansion with Horndeski-like operators at second order in derivatives:
\begin{equation}
\label{def:GMN}
 \mathcal{G}_{\mu\nu}=g_{\mu\nu}+\frac{b_2}{2}\frac{\mpl G_{\mu\nu}}{\Lambda^3}+
 \frac{b_3}{3!}\frac{\mpl}{\Lambda^6} \tilde R_{\mu\alpha\nu\beta}\nabla^\alpha \nabla^\beta \phi+\dots\,,
\end{equation}
where $G_{\mu\nu}$ is the Einstein tensor, $\tilde R_{\mu\alpha\nu\beta}$ the dual Riemann tensor, while $\Lambda$ is the strong coupling scale.  These operators lead to consistent, second order equations of motion. They control the Lorentz-violating, time-dependent scalar background profile that causes cosmic acceleration, as found in eq.~\eqref{eq:scalpr} in the scalar example. They represent the first terms in  the series of higher derivative operators, which will determine the Lorentz invariant, UV completion of the theory~\footnote{Within  the EFT approach, possible pathological  degrees
of freedom introduced by higher-order, higher-derivative operators can be consistently eliminated, by a judicious use of the second-order field equations relative to the lower terms in the perturbative expansion~\cite{Simon:1990ic,Weinberg:2008hq}.}. However, at lower energies  below the scale $\Lambda$, the operators of eq.~\eqref{def:GMN}, together with the scalar background profile, lead to a modification of the speed of GWs. 
\begin{equation}
\label{eq:Hornedski_lf}
\frac{\delta c}{c_{\rm EM}}=\frac{2b_2 \beta^2+b_3 \beta^3}{4+b_2\beta^2+b_3 \beta^3}+\mathcal{O}\left(\frac{k^2}{\Lambda^2}\right)\,,
\end{equation}
where $\beta=\langle \dot \phi \rangle H/\Lambda^3$.  Note that causality considerations do not necessarily demand $\delta c>0$, as this statement is frame dependent~\cite{deRham:2019ctd,deRham:2021fpu}. For simplicity we have considered an example where the EM is minimally coupled to gravity and does not mix with the DE field; but in principle we could imagine generic situations where both EM and GWs differ from perfect luminality at low energies, and recover luminality at higher frequencies within the LVK band. Note however that in practice, EM waves are observed at much higher frequencies than GW ones. 

In what follows we shall consider the cutoff $\Lambda$ of the EFT to lie precisely between LISA and LVK scales, so that at LISA scales, the speed of GWs can be taken to be  $c_T(k\ll \Lambda)\sim c_{\rm EM}-\delta c$, while at LVK scales high-energy effects have to accounted for. At higher energies, higher frequency corrections are relevant and the expansion~\eqref{eq:Hornedski_lf} is no longer valid. Instead a microscopic description takes over, for which the details of the spontaneous breaking of Lorentz invariance at low energy are expected to be irrelevant, and the speed of GWs is expected to be luminal (just like high-frequency light is unaffected by the medium in which it propagates). The transition between the low-frequency regime (for instance between~\eqref{eq:Hornedski_lf}) and the high-energy one are parameterized by the details of the UV completion and can in principle be arbitrarily sharp.

Any modification of the dispersion relation can be already strongly constrained at LVK frequency scales, as proved in many analysis \cite{LIGOScientific:2016sjg,deRham:2016nuf,LIGOScientific:2017bnn,LIGOScientific:2019fpa,LIGOScientific:2020tif,LIGOScientific:2021sio}: see \cite{Harry:2022zey} for the most recent  strongest constraints. We shall therefore take an agnostic approach and focus our attention on the extreme situation where the speed of GWs is considered to be exactly luminal at LVK scales and precisely constant (but non-luminal) at LISA scales, with a sharp transition in between.

In the specific case of  cubic Galileons/Horndeski systems, where
the quantity $\mathcal{G}_{\mu\nu}$ of eq \eqref{defGmn1} reads  $\mathcal{G}_{\mu\nu} = G_3(\phi, (\partial \phi)^2)\Box \phi g_{\mu\nu}$, or
for theories beyond Horndeski,  characterised  for instance by 
$\mathcal{G}_{\mu\nu} = F_4(\phi, (\partial \phi)^2)\epsilon_{\mu\alpha\beta\gamma}\epsilon_{\nu\alpha'\beta'}{}^{\gamma}\nabla^\alpha \nabla^{\alpha'}\phi \nabla^\beta \nabla^{\beta'}\phi$, a resonant decay of luminal GWs
into a subluminal DE field has been pointed out \cite{Creminelli:2019nok}, together with non-linear instabilities that potentially rule out such models at LVK energy scales \cite{Creminelli:2019kjy}. However,
these bounds depend
on the precise details of the EFT and its UV completion. In the present work we focus on different choices of operators, and we additionally rely on a suitable UV completion to ensure that the speed of both GWs and the dark energy field are luminal at LVK frequency scales. While at LISA energy scales the speed of GWs still departs from luminality, at LISA frequencies the decay of gravitational waves into dark energy would be strongly suppressed, as it scales with the fifth power of  the momentum. As for the non-linear instability found in \cite{Creminelli:2019kjy}, it is specific to the cubic Horndeski and beyond Horndeski models; but one may worry that similar non-linear instabilities may also occur for a more generic non-minimal coupling in $\mathcal{G}_{\mu\nu}$, and thus affect our results. However, as indicated in \cite{Creminelli:2019kjy}, the fate of this non-linear instability ultimately depends on the UV completion of the low-energy EFT, and in this work we precisely consider situations for which the UV completion (or partial UV embedding) leads to a decoupling between the dark energy field and GWs at higher energy (as expected from a standard Lorentz-invariant high-energy completion).  

Even in cases where the UV completion may be ineffective at fixing such an instability (e.g. if the latter is occurring at very low energies), it is still possible that the instability is simply due to a breakdown of the Cauchy (initial-data) problem. In fact, 
the instability of \cite{Creminelli:2019kjy}
manifests itself as the effective metric on which the scalar waves travels becoming singular, which is reminiscent of the Tricomi/Keldysh breakdown of the Cauchy problem in K-essence~\cite{Bernard:2019fjb,Bezares:2020wkn,terHaar:2020xxb,Bezares:2021yek,Bezares:2021dma} and non-linear Proca~\cite{Barausse:2022rvg} theory. The latter has been shown to be fixable by suitable choices of gauge and self-interactions~\cite{Bezares:2020wkn,Bezares:2021dma}, and/or by a UV completion~\cite{Bezares:2021yek,Lara:2021piy}.

\begin{figure}
  \centering
    \includegraphics[width=\linewidth]{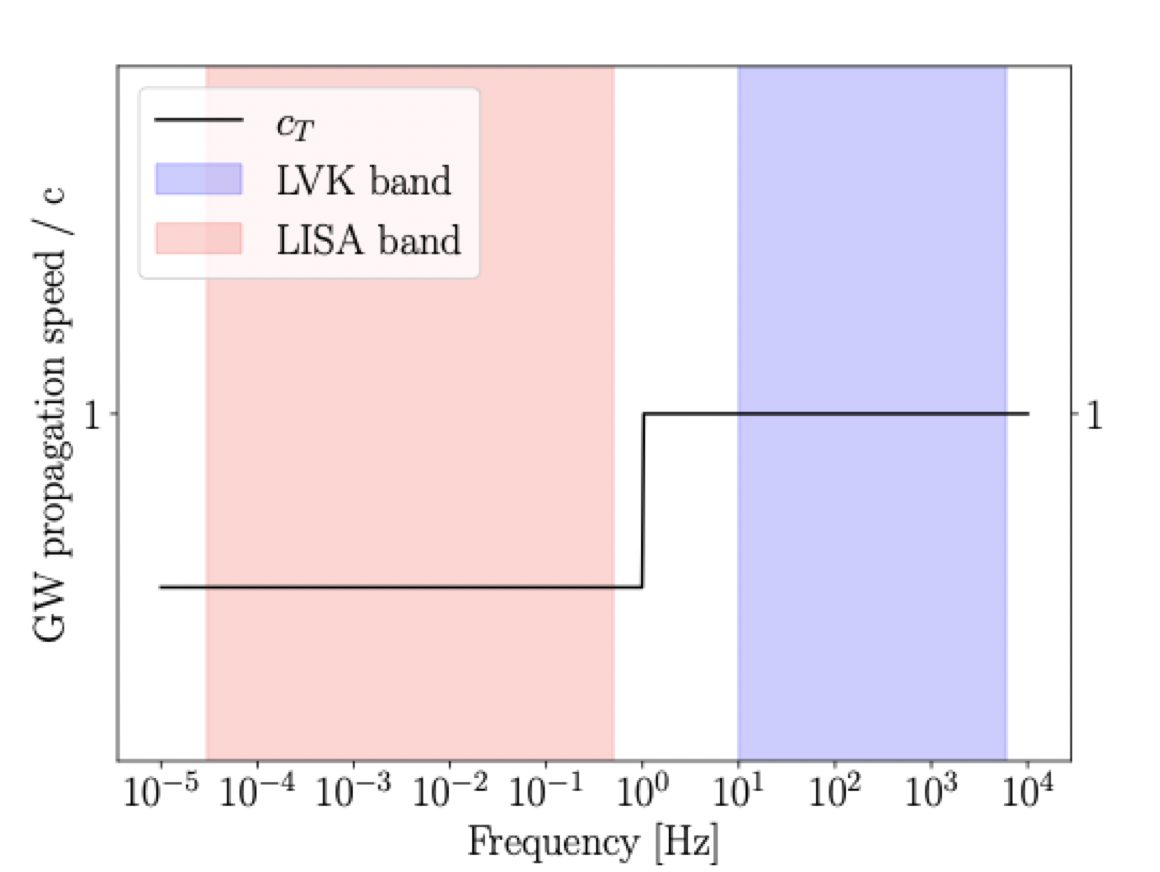}
    \caption{A schematic representation of the frequency-dependent model of $c_T$ considered in this paper.}
    \label{fig:ct}
\end{figure}

\section{GW data analysis and waveform model}
\label{sec:data_analysis}
Having explained the theoretical motivation behind the model we study, in this section we describe the set-up of our forecasts and the parameter sets we consider.

The data $d(t)$ (in time domain) measured by any direct GW wave detector are expressed as a combination of  signal $s(t)$, plus detector noise $n(t)$. The quantity $s(t)$ can be further expanded as the product of the GW signal $h_{ij}(t)$ with the detector response function  $r_{ij}(t)$, which depends on the instrument under consideration: see e.g.~\cite{Romano:2016dpx} for a review on this topic. The interferometer response  to a signal depends both on GW frequency and direction. For  long-duration signals, the detector position  changes over time~\footnote{Typical examples are the  ground-based interferometer motion due to the Earth rotation and revolution, or space-based interferometer motion around the Sun.}. In order to avoid such time-dependence, we split the total observation time $T$ into segments of duration $\Delta T$, with $\Delta T \ll T_{\rm M} $, where $T_{\rm M}$ is the timescale of  the detector motion. In this way,  the detector can be assumed to be at rest within each time segment. Under these assumptions, the data can be expressed in frequency as:
\begin{equation}
\tilde{d}_c(f) = \tilde{n}_c (f) + \tilde{s}_c (f) \; ,
\end{equation}
where the index $c$ runs over the number of  data segments.   The condition of reality  imposes $\tilde{s}_c (f) = \tilde{s}^*_c (-f)$,  and analogously for the noise. For the sake of simplicity, in the following we  drop the $c$ index.\\

Assuming the noise to obey a Gaussian statistics, the log-likelihood for the noise residual can  be expressed as:
\begin{equation}
  -2 \ln \mathcal{L} = \left(  \tilde{d} - \tilde{s}^{\rm th} (f, \vec{\theta}) |  \tilde{d} - \tilde{s}^{\rm th} (f, \vec{\theta})\right) \;  ,
  \end{equation} 
where $\vec{\theta}$ is the vector of parameters  used to characterize the signal, $\tilde{s}^{\rm th} (f, \vec{\theta})$ corresponds to the theoretical model, and $\left( a | b \right)$ is the noise weighted inner product
  \begin{equation}
    (a|b) = 2 \int_{f_1}^{f_2} \frac{a(f)b^*(f)+a^*(f)b(f)}{\tilde{N}(f)} \; \textrm{d} f \; .
\end{equation}
Forecasts on the determination of the model parameters $\vec{\theta}$ can be obtained in terms of the FIM approach~\cite{Finn:1992wt,Cutler:1994ys}. Assuming the minimum of the likelihood matches with the injected value, $\vec{\theta}_0$, the FIM -- corresponding to the second derivative of the loglikelihood evaluated in the best fit -- results 
\begin{equation}
    F_{ij}  = \left. \left(  \frac{\partial \ln \tilde{s}^{\rm th} (f, \vec{\theta}) }{\partial \theta_i  } \right| \left. \frac{\partial \ln \tilde{s}^{\rm th} (f, \vec{\theta}) }{\partial \theta_j  }  \right) \right|_{\vec{\theta} =\vec{\theta}_0 } \;  .
\end{equation}
The signal model, and in particular the GW waveform $\tilde{h}^{\rm th}(f, \vec{\theta})$, can be quite complex depending on the desired level of accuracy. In order to avoid expensive numerical relativity computations, several phenomenological templates, with increasing level of precision, have been introduced~\cite{Ajith:2007kx, Santamaria:2010yb, Husa:2015iqa, Khan:2015jqa, London:2017bcn, Pratten:2020fqn, Garcia-Quiros:2020qpx, Pratten:2020ceb}. Typically, these templates are based on perturbative expansions (for reviews on post-Newtonian theory and on BH perturbation theory see respectively~\cite{Blanchet:2013haa} and~\cite{Kokkotas:1999bd}) to describe the inspiraling and ring-down phases with an intermediate merger region tuned to numerical relativity simulations. In general, the total number of parameters can be quite large, with a minimum of 10 parameters for aligned and non-precessing spins described by a single effective parameter. In the following, and similarly to~\cite{Baker:2022rhh}, we adopt a simplified model where we perform  an average over the sky-localization of the source, as well as over its inclination and polarization. Moreover, we do not include the effects of the spins of the merging bodies. Under  these assumptions, the parameter vector reduces to:
\begin{equation}
    \vec{\theta} = \left\{ \ln \mathcal{M}_o, \ln \eta , d_L, t_c, \Psi_c , \vec{\theta}_{\rm MG} \right\}\,.
\end{equation}
The first five (pure GR) parameters are (the log of) the observed chirp mass of the binary (in solar mass units), (the log of) its symmetric mass ratio, the luminosity distance of the binary, the time to coalescence (in ms) and the coalescence phase. Since we are assuming a sharp transition for the GW speed occurring outside  both of LISA and LVK frequency bands (see Fig~\ref{fig:ct}), $\vec{\theta}_{\rm MG}$  consists of a single parameter $\delta c$, controlling the height of the transition. The exact location of the transition in frequency is irrelevant, as long as it occurs between the LISA and LVK frequency bands, so we do not include it as a parameter.\\

The waveform model adopted in this work is the same as described in~\cite{Baker:2022rhh}, and we refer the reader to this work for details. Our model systematically expresses modifications of amplitude and phase of the waveforms in terms of functions depending on the frequency-dependent GW velocity. For the analysis presented in this paper we consider the lowest order approximation for the amplitude part of the waveform and we expand the phase up to 2.5PN order. Notice though that in the present context we can also exploit the following simplifications:
\begin{itemize}
    \item The transition between the GW speeds  is  sharp, and we assume it to take place outside (and  in between) of  the sensitivity bands of either detectors.  This condition simplifies considerably the expressions for the waveforms discussed in~\cite{Baker:2022rhh}. Relaxing the sharpness of the transition would only lead to more stringent constraints, since a modification of the dispersion relation would then be potentially observable in one of the frequency bands.
     In fact, the transition should be very sharp for ensuring that the GW speed satisfies current bounds at LVK scales:  see Appendix D of \cite{Baker:2022rhh} for a discussion of this topic.
    \item Since the dominant factor in constraining variations in the propagation speed comes from the comparison of  measurements  at the two detectors -- and not from each of the two separate observations --  we further simplify the waveform,  by only considering the common inspiraling part of the signal. We assume here clocks at the LISA and terrestrial detectors are synchronized to high accuracy.
\end{itemize}

\begin{figure}
    \centering
    \includegraphics[width=\linewidth]{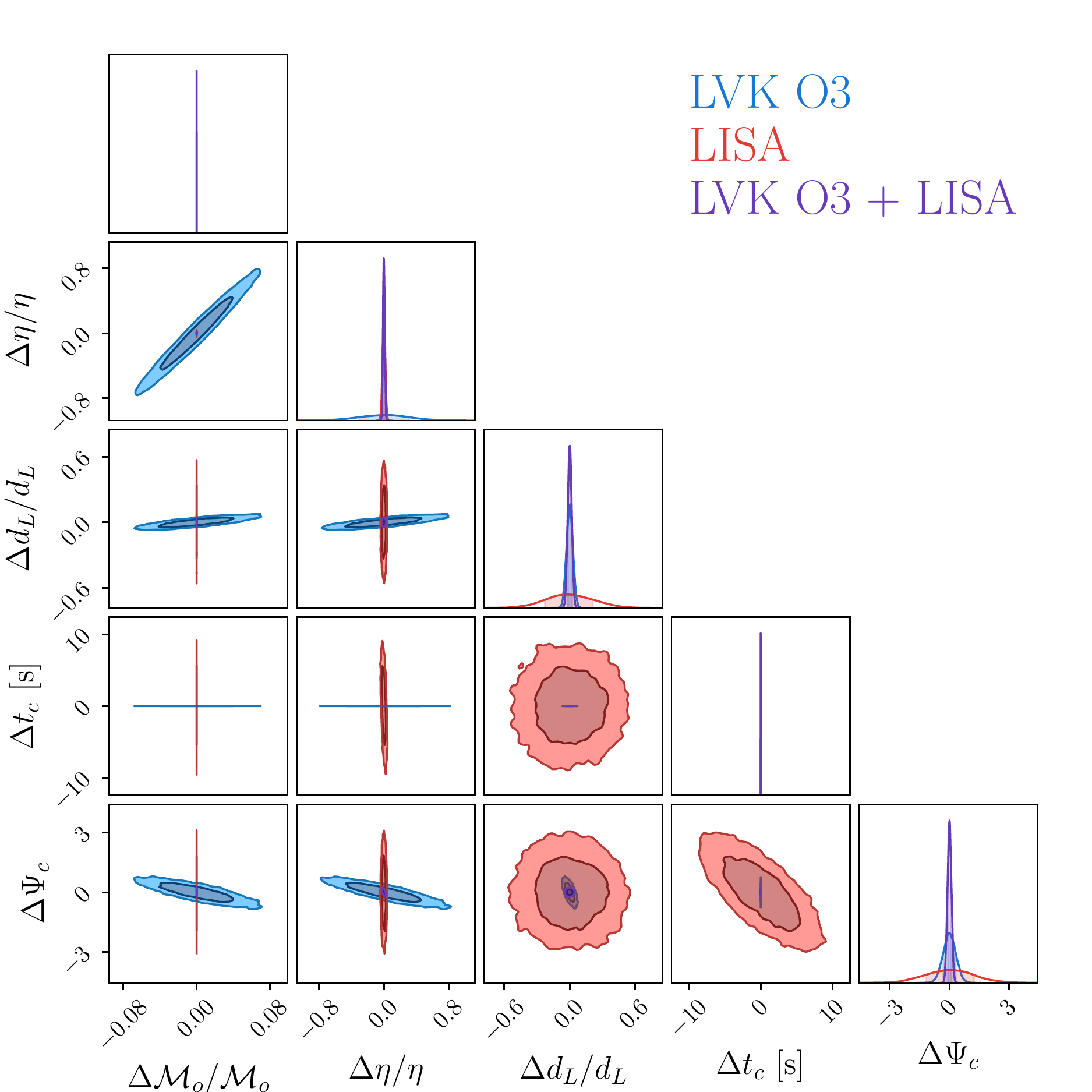}
    \caption{LVK O3 (blue), LISA (red) and joint (purple) constrains for an event with $m_1 = 36 M_{\odot}$, $m_2 = 29 M_{\odot}$, $z = 0.09$, $t_c =0 $, $\Psi_c =0$ assuming pure GR.}
    \label{fig:GR_constraints}
\end{figure}

\begin{figure}
    \centering
    \includegraphics[width=\linewidth]{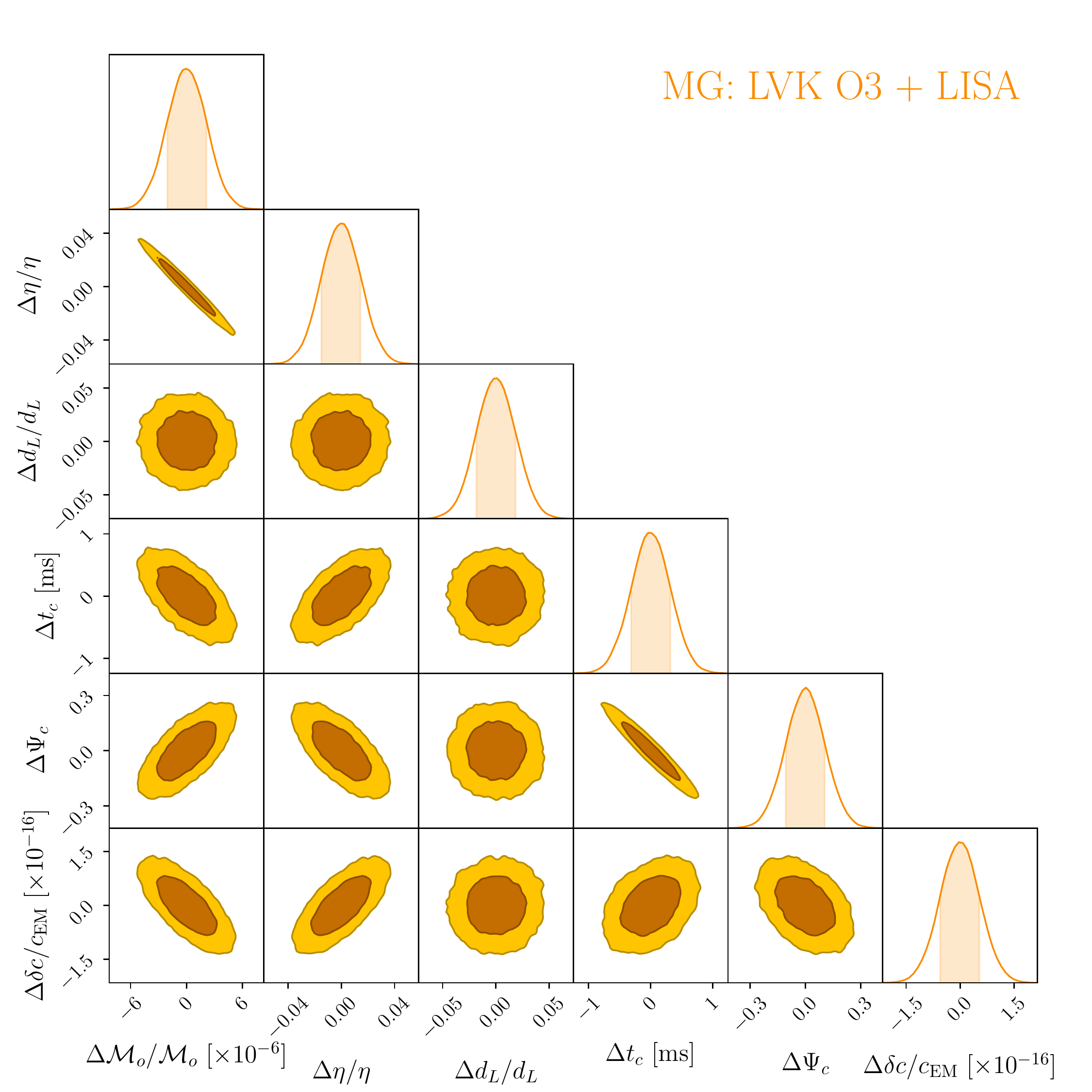}
    \caption{LVK O3 + LISA constrains for an event with $m_1 = 36 M_{\odot}$, $m_2 = 29 M_{\odot}$, $z = 0.09$, $t_c =0 $, $\Psi_c =0$ and $\delta c/c_{\rm EM} = 0$. }
    \label{fig:ct_constraints_GR}
\end{figure}

\begin{figure}
    \centering
    \includegraphics[width=\linewidth]{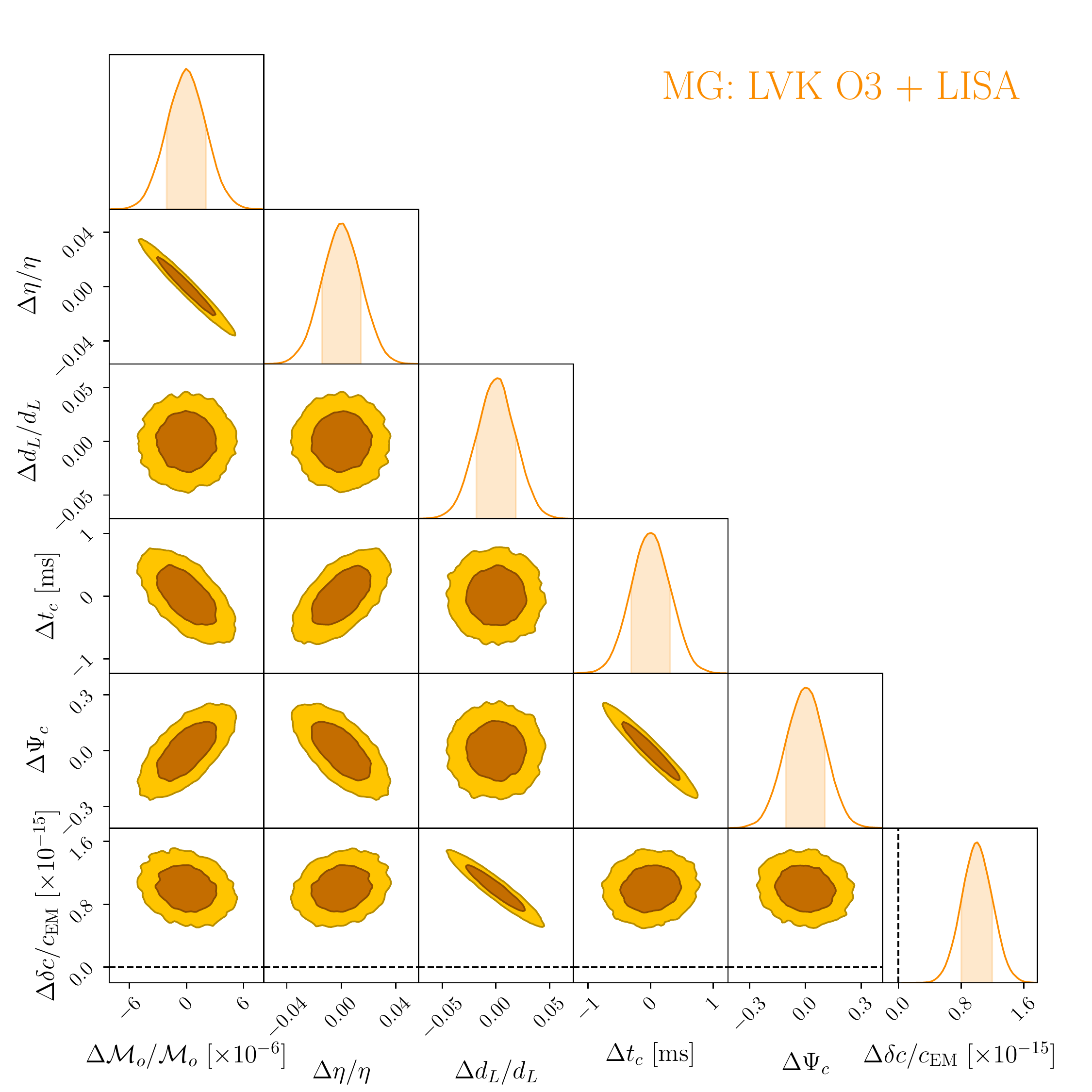}
    \caption{Joint LVK O3 + LISA constrains for an event with $m_1 = 36 M_{\odot}$, $m_2 = 29 M_{\odot}$, $z = 0.09$, $t_c =0 $, $\Psi_c =0$ and $\delta c/c_{\rm EM} = 10^{-15}$.}
    \label{fig:ct_constraints_MG}
\end{figure}

\section{Numerical results and their interpretation}
\label{sec:results}
We   apply   the techniques outlined in  the previous section to an explicit example, aimed at verifying the analytical estimate discussed in Section~\ref{sec:intro}. For this purpose we consider an event with characteristics similar to GW150914, \emph{i.e.} with detector frame masses $m_1 = 36 M_{\odot}$, $m_2 = 29 M_{\odot}$ and at redshift $z = 0.09$. The fiducial values of $t_c$ and $\Psi_c$ are chosen to be zero. For forecasting  LVK constraints, we use  public O3 sensitivity curves for LIGO Livingston, LIGO Hanford, and Virgo\footnote{ 
\url{https://dcc.ligo.org/LIGO-T1500293/public}
.}. For our LISA constraints  we make use of an effective  sensitivity curve~\cite{Flauger:2020qyi}  obtained after averaging    over  sky and  polarisation angle.

As a first step, we provide the constraints that this event imposes on GR parameters. We assume that the minimal detectable frequency is $f_m \simeq 0.017$ Hz, implying that we start to observe the system $\simeq $ 4yrs before the merger takes place. A plot of the constraints obtained from LVK observations, LISA, and  a joint multiband observation of the event are shown in Fig.~\ref{fig:GR_constraints}, while the  specific constraints on  parameters for each case in Fig.~\ref{fig:GR_constraints} are reported  in Table~\ref{table:GR_constraints}. We find that tighter constraints on the chirp mass and the mass ratio are  obtained in the LISA band -- likely because of the long inspiral stage taking place at low frequencies, resulting  in a  phase increase which strongly depends on $\cal{M}$ and $\eta$. However the LVK band provides tighter constraints on the other parameters. This is due to  a higher SNR in the LVK  band, ${\rm SNR}_{\rm LVK}=  55.69$, compared to ${\rm SNR}_{\rm LISA}=  4.7$  in the LISA band. As expected, the joint constraints with both detectors reach a tighter level than each detector individually.

\begin{table}
\centering
\begin{tabular}{||c c c c c c||} 
 \hline
Detector & $\Delta \mathcal{M}_o / \mathcal{M}_o$ & $\Delta \eta / \eta$ & $\Delta d_L /d_L$ & $\Delta t_c [s]$ & $\Delta \Psi_c$ \\ [0.5ex] 
 \hline\hline
 {\rm LVK O3} & 0.027 & 0.31 & 0.029 &  0.00083 &  0.31  \\ 
 \hline
  {\rm LISA} & 2.3 $\times 10^{-6} $ & 0.016 & 0.22 & 3.5 & 1.2 \\
 \hline
 {\rm Combined } &  1.4 $\times 10^{-6} $ & 0.01 & 0.018 &  2.9 $\times 10^{-4}$ & 0.095 \\
 \hline
\end{tabular}
\caption{\label{table:GR_constraints}Table summarizing constraints shown in Fig.~\ref{fig:GR_constraints}.} 
\end{table}
\smallskip

We then proceed by considering models which allow for a sharp transition in $c_T$ to occur at a frequency between the LVK and LISA sensitivity bands. For definiteness,  we choose the transition frequency to be at $1$ Hz. Before studying a non-GR injection, \emph{i.e.} with $\delta c_T / c_{\rm EM} \neq 0$, we first perform a FIM parameter estimation with a non-GR model on a pure GR injection, \emph{i.e.} $\delta c_T / c_{\rm EM} = 0$. This particular example, whose results are shown in Fig.~\ref{fig:ct_constraints_GR}, is considered in order to give a numerical validation to the analytical estimate provided in eq.~\eqref{eq:t_delta_c}. For this event, as we can learn  from the last row of the plot  showing the constraints on $\delta c/c_{\rm EM}$,  deviations from GR are constrained to the $10^{-16}$ level. Besides constraints on the parameter $\delta c/c_{\rm EM}$, Fig.~\ref{fig:ct_constraints_GR} also shows the forecast constraints on ${\cal M}_o$, $\eta$, $d_L$ and $\Psi_c$. As expected, given that an additional parameter is involved in the model,  the constraints slightly worsen with respect to the pure GR case summarized in the last line of Table~\ref{table:GR_constraints}. However, their size keeps  the same order of magnitude as in the GR example.

The final case we consider in this work is a non-GR injection. In particular, we will consider the case where the height of the transition taking place at $f = 1 $Hz is $\delta c/c_{\rm EM} = 10^{-15}$. A plot of the forecast constraints of this event for a joint observation with the LVK detectors and LISA are shown in Fig.~\ref{fig:ct_constraints_MG}. In this plot, the values corresponding to  GR (\emph{i.e.} $\delta c/c_{\rm EM} =0$) are shown as a dashed black line in the bottom row of plots. We find for the $1\sigma$ bound of $\delta c/c_{\rm EM}$ the value $1.93 \times10^{-16}$, which means that our fiducial value $\delta c/c_{\rm EM} = 10^{-15}$ can be comfortably detected, and the GR  value $\delta c=0$  excluded at more than $5 \sigma$ level.  As for the pure GR injection, the forecast constraints on all the other parameters are only mildly affected by the introduction of the additional parameter. 

 As described in~\cite{Harry:2022zey}, a deviation of $\delta c/c_{\rm EM} = 3\times10^{-15}$ would be sufficient to shift by 2 minutes the coalescence time $t_c$ in the LVK band. Given our FIM forecast on $t_c$ for GW150914-like events, in such a case we will be almost certain to detect the time shift with respect to the LISA prediction. This is compatible with the results in~\cite{Harry:2022zey} obtained using a time-shift method.
    
Notice also that if the transition in $c_T(f)$ occurs  within the LISA band, constraints on $\delta c$ can be obtained with the GW signals from binary massive black hole mergers~\cite{Baker:2022rhh}. However, if the sharp transition takes place at frequencies lower  than the LISA band, the joint detection with LIGO-Virgo-KAGRA and LISA will not be able to detect deviations on  $c_T$, since $\delta c=0$ within their sensitivity bands. Such a low-scale transition would occur in scenarios with a parametrically very   low cutoff scale as could be motivated from high-energy bounds. Constraints on low-energy EFTs coming from consistency of the high energy completion (in particular improved versions of positivity bounds~\cite{Bellazzini:2017fep,deRham:2017xox} and recently derived fully crossing symmetric positivity bounds~\cite{Tolley:2020gtv,Caron-Huot:2020cmc,Sinha:2020win}) indicate that the cutoff in theories related  to~\eqref{def:GMN} is often required to be much lower than the expected strong coupling scale $\Lambda$.

To summarise, the forecasts performed here -- using a 2.5PN order waveform -- indicate that fractional changes to the speed of GWs as small as $10^{-15}$ can be cleanly detected with a single multiband GW system. Furthermore, the remaining `standard' parameters of the system are not degraded in any significant way, allowing for reliable analysis in a multiparameter space.

\section{Conclusions}
\label{sec:conclusions}
The propagation of GWs provides an exquisite probe into the nature of gravity and its couplings with the dark sector (see~\cite{Lagos:2019kds,Mastrogiovanni:2020gua,Baker:2020apq,Finke:2021aom,Finke:2021znb,Finke:2021eio,Mancarella:2021ecn,Ezquiaga:2021ler,Ezquiaga:2021ayr,Leyde:2022orh,Ezquiaga:2022nak,Caliskan:2022hbu} for other GW propagation-related work). Motivated by the late-time acceleration of the Universe, EFTs of DE and infrared EFTs of gravity have emerged as systematic ways to parameterize couplings between gravity and the fields driving the acceleration. In the regime of validity of these EFTs (or in their weakly coupled regions), the speed of GWs can typically differ from that of the EM counterpart. However at higher frequencies,  the degrees of freedom driving the late-time acceleration of the Universe are expected to be insignificant, and hence their effects on the propagation of GWs are potentially undetectable by observations in the LVK band alone. 

In this work we explored the possibility that the propagation of GWs undergoes a sharp transition between the LISA and the LVK frequency bands. We proved that a single  SOBHB merger event detected in both bands would be sufficient to either confirm a change in speed or severely constrain any scale-dependent transition between $10^{-4}$ and $10^4$Hz. This conclusion can be reached analytically, but   we have corroborated it by a FIM analysis of  future joint LISA-LVK observations. Our analysis confirms that the observation of a single event in the both the LVK and LISA bands would allow for constraining the speed of GWs within the LISA band at the  $10^{-15}$ level (at $\sim 5 \sigma$).

Such a constraint would strongly reduce the allowed region of parameter space of the EFT of DE, and further limit models of modified gravity with modified dispersion relations. Besides testing the nature of DE (and its coupling with gravity), a multiband constraint on the propagation of GWs would provide a significant new insight on the nature of gravity in the low-energy regime and its couplings (self-interactions or couplings with other light degrees of freedom). In particular it would be interesting to explore how birefringence effects highlighted in~\cite{Akrami:2018yjz} would further constrain model of modified gravity.

The rate of heavy SOBHBs detectable in the LISA band remains highly uncertain at present~\cite{Tamanini:2016zlh}. However, the results laid out here indicate that a single multiband system will have a significant impact on the field,  analogous to the first multimessenger event and settling questions raised in its wake.\\

\noindent\textit{Acknowledgments:} 
We are grateful to Johannes Noller and Ian Harry for useful discussions on this topic and their related work, and for  useful feedback on our draft.

T.B. is supported by ERC Starting Grant \textit{SHADE} (grant no.~StG 949572) and a Royal Society University Research Fellowship (grant no.~URF$\backslash$R1$\backslash$180009). E.B. acknowledges support from the European Union's H2020 ERC Consolidator Grant ``GRavity from Astrophysical to Microscopic Scales'' (Grant No.  GRAMS-815673) and the EU Horizon 2020 Research and Innovation Programme under the Marie Sklodowska-Curie Grant Agreement No. 101007855. A.C. is supported by a PhD grant from the Chinese Scholarship Council (grant no.202008060014).
CdR acknowledges support from a Wolfson Research Merit Award, the Simons Foundation award ID 555326 under the Simons Foundation Origins of the Universe initiative, Cosmology Beyond Einstein's Theory and the Simons Investigator award 690508. CdR and M.P. are also funded by STFC grants ST/P000762/1 and ST/T000791/1 as well as by the European Union’s Horizon 2020 Research Council grant 724659 MassiveCosmo ERC-2016-COG. G.T. is partially funded by the STFC grant ST/T000813/1. 
 For the purpose of open access, the authors have applied a Creative Commons Attribution licence to any Author Accepted Manuscript version arising.

\begin{appendix}

\section{The constraint on $c_T$}
\label{app_const}

In this Appendix we expand on the  arguments outlined in  the main text, relating  the difference $\delta c\,=\,c_T-c_{\rm EM}$ with the difference $|t_{\rm Lisa }-t_{\rm Earth}|$  between the `expected' (by LISA) and `measured' (by a ground-based detector) times of coalescence.   During a binary inspiral and merging process, the change in the orbital radius renders the GW frequency time-dependent. At leading order in a post-Newtonian (PN) expansion, we find the relation  (see e.g.~\cite{Baker:2022rhh})
 \begin{equation}
 \label{eqfrder1}
\frac{ d f}{d t_{\rm ret}}\,=\,\frac{96}{5} \,\pi^{8/3}\,{\cal M}^{5/3} \,{\cal F}\left(f \right)\,f^{11/3}\,,
 \end{equation}
 with $f$ the GW frequency measured at the detector position. $t_{\rm ret}$ indicates  retarded time
 \begin{equation}
 \label{def_tret}
 t_{\rm ret}\,=\,t-\frac{D}{c_T(f)}\,,
\end{equation} 
with $t$  the time measured at detector position, 
  $D\,=\,a(t) r$  the  distance  of the merging binary ($r$  being the comoving distance, and $a(t)$ the scale factor) and ${\cal M}$ is the binary chirp mass. The  function ${\cal F}$  depends on the GW speed $c_T$, and is   assumed to be itself a function of frequency. The structure of ${\cal F}$ depends
 on the scenarios and approximations considered: see for example~\cite{Baker:2022rhh} for more details and explicit examples; {in the limit of GR, ${\cal F}=1$}. In deriving equations~\eqref{eqfrder1} and~\eqref{def_tret}, we  work in a nearly-static regime where the explicit time-dependence of the scale factor  can be neglected, as appropriate for low-redshift sources. Moreover, we assume that $\delta c/c$ is extremely small, so that in first approximation $c_T$ does not influence the amplitude of the waveform. 
 
We can re-arrange and integrate both sides of eq.~\eqref{eqfrder1}, and obtain
\begin{widetext}
\begin{equation}
\frac{5}{96\,\pi^{8/3}\,{\cal M}^{5/3} }\,\int_{f_c}^f\,\frac{d \hat f}{\hat f^{11/3}} + \frac{5}{96\,\pi^{8/3}\,{\cal M}^{5/3}}\,\int_{f_c}^f\,\frac{d \hat f}{\hat f^{11/3}}\left(\frac{1}{{\cal F}(\hat f)}-1\right)\,=\,(t-t_c)+\left(\frac{D}{c_T(f_c)}
- \frac{D}{c_T(f)}\right)\,.
  \end{equation}
\end{widetext}
If $c_T$ were one, the second terms in both sides of this equation would vanish. One would find the standard GR  expression for the time $\tau^{\rm GR}\,=\,t-t_c$ to coalescence:
\begin{equation}
\tau^{\rm GR}\,=\,\frac{5}{96\,\pi^{8/3}\,{\cal M}^{5/3} }\,\int_{f_c}^f\,\frac{d \hat f}{\hat f^{11/3}}
\,.
\end{equation}
 Assuming standard GR formulas, LISA observations would then  estimate the coalescence time 
$t_{\rm LISA}$. 
Instead,  the actual detection of the coalescence process with ground-based experiments would measure the effects  of  a frequency-dependent $c_T$ (if present), and a coalescence time $t_{\rm Earth}$ different from what predicted in GR. Taking the difference among the two cases, we find
\begin{widetext}
  \begin{equation}
t_{\rm LISA}-t_{\rm Earth}\,=\,\frac{D}{c_{\rm EM}}\,\left(
\frac{c_{\rm EM}-c_T(f)}{c_T(f)}\right)+{\tau^{\rm GR}}\,
\left[
\int_{f_c}^f\,\frac{d \hat f}{\hat f^{11/3}}\left(\frac{1}{{\cal F}(\hat f)}-1\right)
 \Big/\left( \int_{f_c}^f\,\frac{d \tilde f}{\tilde f^{11/3}}\right)\right]\,,
\label{genforA1}
  \end{equation}
  \end{widetext}
where, since coalesce occurs at ground-based detector frequencies, we assume $c_T(f_c)\,=\,c_{\rm EM}$. Starting from  eq.~\eqref{genforA1}, we can  estimate  the order of  magnitude for the bounds on $\delta c$ offered by multiband measurements. We expect that the uncertainties on measurements of the coalescence time are of order ${\cal O}(1)$ s.  Considering  an event as GW150914, occurring at a distance of order $D\,=\,410$ Mpc, and assuming it is first detected in the LISA frequency band, we have a time to coalesce  of order $\tau^{\rm GR}\,=\,5$ years. Hence, $D/c_{\rm EM}\,\sim\,4.2\times 10^{16}$ s, while $\tau^{\rm GR}\,\simeq\,1.6 \times 10^8$ s.  Making the hypothesis that the contribution within square parenthesis in eq.~\eqref{genforA1} is at most of order one -- as expected for physically motivated choices of ${\cal F}$, see e.g.~\cite{Baker:2022rhh} -- the second term of eq.~\eqref{genforA1} is orders of magnitude smaller, and can be neglected with respect to the first.  Bounds on $\delta c\,=\,|c_T-c_{\rm EM}|$ reach values of
 \begin{equation}
 \frac{\delta c}{c_{\rm EM}}\,\le\,\frac{1\, {\rm s}}{4.2\times 10^{16} \,{\rm s}}\,\simeq\,10^{-16}\,,
 \end{equation}
 as stated in the main text.

\end{appendix}
\bibliographystyle{utphys}
\bibliography{master}
\end{document}